\documentclass[preprint,12pt]{elsarticle}

\usepackage{amssymb}
\usepackage{comment}
\usepackage{url}
\usepackage{graphicx}
\usepackage{caption}
\usepackage{subcaption}
\usepackage{color, colortbl}
\usepackage{tabularx}
\usepackage{booktabs}
\usepackage{multirow}
\usepackage{hyperref}

\definecolor{gray}{gray}{0.9}
\definecolor{lightblue}{rgb}{0.88,1,1}
\newcolumntype{a}{>{\columncolor{gray}}c}

\newcommand{\cc}[1]{{channels}}

\newcommand{\RqOne}{PS$_{1}$: Are we able to distinguish knowledge within the communication channels of GitHub projects?}

\newcommand{\RqTwo}{RQ$_{1}$: Do communication channels change over time?}

\newcommand{\RqThree}{RQ$_2$: Do communication channels differ within ecosystems?}

\journal{The Journal of Systems and Software}

\begin{document}

\begin{frontmatter}

\title{A Topological Analysis of Communication Channels for Knowledge Sharing in Contemporary GitHub Projects}

\author[label1]{Jirateep Tantisuwankul}
\author[label2]{Yusuf Sulistyo Nugroho}
\author[label2]{Raula Gaikovina Kula}
\author[label2]{Hideaki Hata}
\author[label1]{Arnon Rungsawang}
\author[label1]{Pattara Leelaprute}
\author[label2]{Kenichi Matsumoto}
\address[label2]{Nara Institute of Science and Technology, Japan}
\address[label1]{Kasetsart University, Thailand}

\begin{abstract}
With over 28 million developers, success of the GitHub collaborative platform is highlighted through an abundance of communication \cc{} among contemporary software projects. 
Knowledge is broken into two forms and its sharing (through \textit{communication channels}) can be described as externalization or combination by the SECI model.
Such platforms have revolutionized the way developers work, introducing new channels to share knowledge in the form of pull requests, issues and wikis. 
It is unclear how these channels capture and share knowledge.
In this research, our goal is to analyze these communication channels in GitHub.
First, using the SECI model, we are able to map how knowledge is shared through the communication channels.
Then in a large-scale  topology analysis  of seven library package projects (i.e., involving over 70 thousand projects), we extracted insights of the different communication channels within GitHub. 
Using two research questions, we explored the evolution of the channels and adoption of channels by both popular and unpopular library package projects. 
Results show that (i) contemporary GitHub Projects tend to adopt multiple communication channels, (ii) communication channels change over time and (iii) communication channels are used to both capture new knowledge (i.e.,  externalization) and  updating existing knowledge (i.e., combination).
\end{abstract}



\end{frontmatter}

\section{Introduction}
\label{sec:introduction}
A key ingredient to the emergence and success of software projects on collaborative platforms such as GitHub\footnote{\label{githubweb}\url{https://github.com/}} has been its distributed information sharing nature, which is the ability to interact and share information between software developers.
With over 28 million developers and 67 million repositories reported in 2018, GitHub hosts a multitude of diverse developer ecosystem (also referred to as communities)\footnote{The survey result is available at \url{https://octoverse.github.com/}}.
As well as hosting traditional software projects, GitHub is also the home to sometimes trivial library projects \cite{Abdalkareem:2017} and have been the focus of recent studies \cite{Mens:2016,Kula:2018,Mirhosseini:2017,Gonzalez-Barahona:2017,Hejderup:2018,Kikas:2017,Decan:2018}. 
For instance, package managers like npm\footnote{website at \url{https://www.npmjs.com/}} host around 700 thousand packages on GitHub.
Interestingly, we find that a single npm developer could be the maintainer for hundreds of these packages.

Investing in knowledge creates value during software development, especially in the context of human capital~\cite{DBLP:journals/corr/abs-1805-03844}.
This knowledge can then be represented and shared in contemporary software through social and technical \textit{`communication channels'}, mostly used to improve and maintain a project's presence in an ecosystem.
Examples of such channels include forking, pull requests, the readme file documentation and so on.
According to the business management perspective, communication \cc{} can be distinguished into either tacit and explicit knowledge, with its transfer being described as externalization and combination (using the SECI model~\cite{nonaka1995knowledge}).
In fact, open source projects heavily rely on social markers and its popularity (i.e., star counts and forks) for measuring their abilities to attract and maintain their contributors.
Although much work has covered the different communication channels, a mapping of all these communication channels and the knowledge shared has not yet been studied.

The research gap that this paper fills is understanding at the topological level of how knowledge is shared between communication channels of contemporary projects.
There has been work that has studied the social collaborations between projects, but not an analysis of multi-channels over large-scale ecosystem of libraries.
A study by Storey et al.~\cite{Storey17} showed that communities of FLOSS (Free Libre Open Source Software) projects are shaped through social and communication channels (also referred to as social coding).
Recently, Aniche et al.~\cite{Aniche:2018} confirmed that news channels also play an important role in shaping and sharing knowledge among developers.

In this paper, we investigate communication channels to understand how projects share knowledge at the software ecosystem level.
Inspired by the knowledge-based theory of the firm~\cite{doi:10.1002/smj.4250171110}, our study is to validate the underlying theory behind the transferable of knowledge within these library ecosystems, and to investigate how ecosystems influence social practices within and outside their ecosystems.
To achieve our goal that is to analyze how communication channels share knowledge over projects, we first identify different knowledge forms of channels in over 210 thousand library projects from seven different library ecosystems. 
We then explore the evolution of these \cc{} and distinguish differences between these seven ecosystems.
Similar to a study by Lertwittayatrai et al.~\cite{Lertwittayatrai:2017:1710.00446}, we use topological data analysis to generate topologies that cover three years (i.e., 2015 to 2017). 
Using topology data analysis, the results of the study show that (i) contemporary GitHub Projects tend to adopt multiple communication channels, (ii) communication channels change over time, and (iii) communication channels are used to capture new knowledge (i.e., externalization) and  updating existing knowledge.
The contributions of the study are two-fold. 
First, we present a manual categorization of \cc{} forms in software projects. 
The second contribution is a large-scale analysis of \cc{} for software projects over seven ecosystems using the topological analysis of software library projects for seven different software ecosystems.

The rest of the paper is organized as follows. 
Section~\ref{preliminaries} describes our initial work to classify the communication \cc{} into tacit or explicit. 
Section~\ref{sec:TMS} details our experiments using the topological data analysis of the seven GitHub library ecosystems.
Section \ref{sec:evaluation} is the evaluation of our topological data analysis technique.
Section~\ref{sec:implications} discuss the implication of the experimental results, with Section \ref{sec:threats_to_validity} defining the threats of validity. 
We present the related works in Section \ref{sec:developer_knowledge_sharing}, and finally conclude the papers and summarize potential avenues for future work in Section \ref{sec:conclusion}.
The replication package that contains all the dataset and experiment details are accessible from \url{https://github.com/NAIST-SE/TDA_Communication_Channels}.

\section{Preliminary Study: Communication Channels and Knowledge Sharing in Software Projects}
\label{preliminaries}
Before we proceed with the study, we first carried out a preliminary study to first understand what knowledge exists and is transferred through the communication channels.

\subsection{Motivation}
The study of knowledge sharing has had an impact in fields like Sharing Architectural Knowledge~\cite{ZAHEDI2016995}, where architectural  decision-making  and  has  been shown to increase project consistency,  coordination,  and communication coherence over time.
To understand knowledge sharing, we apply and distinguish different knowledge forms to communication channels.
We use existing models of knowledge and how they are transferred.
We carry out an empirical study to analyze and answer the formulated research question: \textit{\textbf{\RqOne}}

\subsection{Approach}
Our approach to answer the preliminary study question is through analysis of historical information. 
We first carried-out an investigation of possible communication channels, which we then methodologically classify into the different knowledge forms.
We then use the SECI model to understand the knowledge transfer within these \cc{}.

As shown in Table \ref{tab:tacitandexplicitknowledge}, there are two knowledge forms \cite{polanyi_1962,polanyi_1966, nonaka1995knowledge}. The first is tacit knowledge (know-how) where the knowledge is embedded in the human mind through experience and jobs. Personal wisdom and experience, context-specific are more difficult to extract and codify. 
In addition, tacit knowledge includes insights and intuitions.
The second is explicit knowledge (know-that) which is codified and digitized in books, documents, reports, memos, etc. This type of knowledge is easily identified, articulated, shared and employed that can facilitate action.
To classify the transfer (through sharing) of knowledge within each communication channel, we used the SECI knowledge model.
Nonaka and Takeuchi's SECI model is a model of knowledge dimensions that describes the transformation of tacit and explicit knowledge into organizational knowledge~\cite{nonaka1995knowledge}.
Since it was first introduced by Nonaka~\cite{Nonaka:1990}, SECI model has been used in many area of studies.
D{\'a}videkov{\'a} et al.~\cite{Davidekov:10.1007/978-3-319-50337-0_9} used SECI model to analyze various information and communication technology (ICT) tools in bridging virtual collaboration between team members without their physical presence.
In comparison with traditional teams that requires the presence of individuals, virtual collaboration demands the motivation of team members, support from team leader, and appropriate technology.
Therefore, the preference of such suitable ICT tools for each activity in organizations is necessary.
As shown in Table~\ref{tab:secimodel}, SECI model contains four dimensions of knowledge which together form the acronym ``SECI''.
In this paper, we focus specifically on the \textit{externalization} and \textit{combination} in our classifications.

\begin{table}[]
    \caption[Caption for LOF]{Distinctions between Tacit and Explicit Knowledge\footnotemark}
    \label{tab:tacitandexplicitknowledge}
    \centering
    \footnotesize
    \resizebox{\columnwidth}{!}{%
    \begin{tabular}{|l|l|}
        \hline\noalign{\smallskip}
        \textbf{Tacit Knowledge} & \textbf{Explicit Knowledge}  \\
        \hline\noalign{\smallskip}
        T1{:} Subjective, cognitive, experiential learning & E1{:} Objective, rational, technical   \\
        \rowcolor{gray}
        T2{:} Personal & E2{:} Structured   \\
        \rowcolor{gray}
        T3{:} Context sensitive/specific & E3{:} Fixed content  \\
        \rowcolor{gray}
        T4{:} Dynamically created & E4{:} Context independent   \\
        T5{:} Internalized & E5{:} Externalized \\
        T6{:} Difficult to capture and codify & E6{:} Easily documented \\
        T7{:} Difficult to share & E7{:} Easy to codify \\
        T8{:} Has high value & E8{:} Easy to share  \\
        T9{:} Hard to document & E9{:} Easily to transferred/taught/learned \\
        T10{:} Hard to transfer/teach/learn & E10{:} Exists in high volumes   \\
        T11{:} Involves a lot of human interpretation &    \\
        \hline
    \end{tabular}%
    }
\end{table}
\footnotetext{\url{https://www.tlu.ee/~sirvir/Information\%20and\%20Knowledge\%20Management/Key_Concepts_of_IKM/tacit_and_explicit_knowledge.html}}

\begin{table*}[]
    \centering
    \caption{Taken from Nonaka and Takeuchi~\cite{nonaka1995knowledge}, four dimensions of knowledge transfer}
    \label{tab:secimodel}
    \resizebox{\textwidth}{!}{
    \begin{tabular}{llp{8cm}}
        \toprule
        \textbf{Dimension} & \textbf{Knowledge Transfer} & \textbf{Description}  \\
        \midrule
        Socialization & Tacit to Tacit & Social interaction as tacit to tacit knowledge transfer   \\
        \rowcolor{gray}
        Externalization & Tacit to Explicit & Articulating tacit knowledge through dialogue and reflection. When tacit knowledge is made explicit, knowledge is crystallized, thus allowing it to be shared by others, and it becomes the basis of new knowledge   \\
        \rowcolor{gray}
        Combination & Explicit to Explicit & Systemizing and applying explicit knowledge and information    \\
        Internalization & Explicit to Tacit & Learning and acquiring new tacit knowledge in practice   \\
        \bottomrule
    \end{tabular}
    }
\end{table*}

The identification of knowledge within communication channels was performed by a group consensus among three of the authors, with rationale clearly aligned with the formal definitions.

\subsection{Data Collection}
\label{sec:dataset}
For the preliminary study, the authors used the \texttt{libraries.io}\footnote{\url{https://libraries.io/data}} collection of GitHub software projects.
This dataset includes various communication \cc{} and covers the largest range of ecosystems.
According to its website, libraries.io indexes data from over 3 million library packages from 36 package managers. 
Package managers represent different ecosystems of libraries.
For example, libraries belonging to the nodeJS package manager (npm) are part of the bigger JavaScript ecosystem of projects.
Furthermore, libraries.io also monitors and stores package releases, analyzes each project's code, ecosystem, distribution and documentation, and map the relationships between packages.
Our dataset has also been used in recent empirical studies \cite{Kikas:2017,Decan:2018}.

\begin{table}[]
    \caption{Seven Library Package Platform Ecosystems}
    \begin{center}
        \resizebox{\columnwidth}{!}{%
        \begin{tabular}{llp{4.5cm}rrrr}
            \hline
            \textbf{Library} & \textbf{Programming} & \textbf{Typical Usage} & \multicolumn{4}{c}{\textbf{\# Stars}}\\
            \noalign{\smallskip}
            \cline{4-7}
            \textbf{Package Manager} & \textbf{Language} & \textbf{Domain} & \textbf{Min} & \textbf{Max} & \textbf{Median} & \textbf{Mean} \\
            \hline
            Go & GoLang & Developed by Google Applications & 592 & 92,227 & 6,866.24 & 2,559\\
            npm & nodeJS JavaScript & Web Services & 569 & 122,630 & 8,479.49 & 3,372\\
            Packagist & PHP & Server-side web development & 8 & 122,630 & 194.13 & 23\\
            RubyGems & Ruby & Web Applications & 11 & 90,383 & 433.96 & 48\\
            PyPI & Python & General scripting & 10 & 122,630 & 439.77 & 39\\
            Bower & JavaScript & Web Services & 5 & 122,630 & 866.11 & 43\\
            Maven & Java-based languages & Languages that use Java Virtual Machine & 107 & 122,630 & 1,755.45 & 454\\
            \hline
        \end{tabular}%
        }
    \label{tab:library_platform_names}
    \end{center}
\end{table}

As shown in Table \ref{tab:library_platform_names}, our collected raw dataset is a subset of the seven largest library ecosystems from the \texttt{libraries.io} dataset.
Furthermore, we used the star count to as to get the more popular repositories within each ecosystem~\cite{Borges:7816479}. 
The higher star ensures that the package has value to the ecosystem.
Thus, the top 10,000 ranked projects from each ecosystem was collected.
Two authors then identified and mapped 13 communication channels from the raw dataset features. 
Details of the mapping are discussed in the replication package and presented in Table~\ref{tab:desc_of_feature}.

Each library ecosystem is described below.
Go\footnote{\url{https://golang.org/}} is a package manager in GoLang programming language which is developed by Google. The npm\footnote{\url{https://www.npmjs.com/}} and Bower\footnote{\url{https://bower.io/}} which are renowned for the JavaScript are mostly used in the website development. Similar to the npm and Bower,
Packagist\footnote{\url{https://packagist.org/}} is very common for the website development but in server-side. The language used for this package is PHP. Meanwhile, RubyGems\footnote{\url{https://rubygems.org/}} is a framework of library management contains functions that can be called by a Ruby program. Finally, the python-based library manager, PyPI\footnote{\url{https://pypi.org/}} works for writing script in general, while the Java-based language that use Java Virtual Machine (JVM) is Maven\footnote{\url{https://maven.apache.org/}}.

\subsection{Analysis}
\underline{{Answering PS$_{1}$}}: Using the collected dataset, we labeled each of communication channel to a knowledge form (i.e., tacit or explicit).
The manual labeling was performed by one author and later validated by other co-authors.
Based on Table \ref{tab:tacitandexplicitknowledge}, we found that labeling T2, T3, T4, E2, E3 and E4 were the most identifiable distinctions. 
To reduce bias, the first author and second author did independent labeling. 
Then, in a round table, other authors were consulted for any conflicts.
In Table \ref{tab:desc_of_feature}, we provide a full rationale for each feature.

\subsection{Results}
We now present the results of classifying knowledge of each channel.
\subsubsection{\textbf{\RqOne}}
\begin{quote}
\textit{Yes, we are able to distinguish knowledge forms \cc{} in software projects. }
\end{quote}

\begin{table*}[t]
    \caption{Summary of 13 channels classified with rationale.}
    \begin{center}
        \resizebox{\columnwidth}{!}{%
        \begin{tabular}{|c|l|l|}
            \hline
            & & \textbf{Coding (Table~\ref{tab:tacitandexplicitknowledge})}    \\
            \textbf{Dimensions} & \textbf{Channels} & \textbf{Rationale}  \\
            & & \textbf{Source} \\
            \hline
            Externalization & GitHub Pages & T2, T3 \\
                & & Personal webpage of a project, the content is specific, and it has no standard template to create. \\ 
                & & \url{https://help.github.com/en/articles/what-is-github-pages} \\
                \cline{2-3}
            & Readme & T3, T4 \\
                & & The content is specific and is created dynamically without a template. \\
                & & \url{https://help.github.com/en/articles/about-readmes} \\
                \cline{2-3}
            & Security Audit & T2, E3 \\
                & & Although the audit is personal, the contents are fixed. \\
                & & \url{https://help.github.com/en/articles/reviewing-the-audit-log-for-your-organization} \\
                \cline{2-3}
            & Wiki & T2, T3 \\
                & & Similar to GitHub Pages, the contents of wiki are personal and specific. It has no specific template to create. \\
                & & \url{https://help.github.com/en/articles/about-wikis} \\
            \hline
            Combination & Changelog & E2, E3 \\
                & & The changes are documented in a structured manner, the contents are fix and cannot be customized. \\ 
                & & \url{https://github.blog/2018-05-03-introducing-the-github-changelog/} \\
                \cline{2-3}
            & Code of Conduct & E2, E3 \\
                & & There is a standard template to make the contents of code of conduct. \\ 
                & & \url{https://help.github.com/en/articles/adding-a-code-of-conduct-to-your-project} \\
                \cline{2-3}
            & Contributing & E2, E3, E4 \\
                & Guidelines & The contents are structured, fixed and independent. It is created by following a template. \\
                & & \url{https://help.github.com/en/articles/setting-guidelines-for-repository-contributors} \\
                \cline{2-3}
            & Fork & E2, E3, E4  \\
                & & Fork has structured and fixed content. The context is independent.  \\
                & & \url{https://help.github.com/en/articles/about-forks} \\
                \cline{2-3}
            & Issue Tracker & E2, E4 \\
                & & The contents are independent and adopted from a system in a structured way. \\
                & & \url{https://en.wikipedia.org/wiki/Issue_tracking_system} \\
                \cline{2-3}
            & License & E2, E3 \\
                & & The contents of license are structured and fixed. \\
                & & \url{https://help.github.com/en/articles/licensing-a-repository} \\
                \cline{2-3}
            & Security Threat & E2, E3, E4 \\
                & Model & The security regulations that are structured, fixed and independent.  \\
                & & \url{http://www.agilemodeling.com/artifacts/securityThreatModel.htm} \\
                \cline{2-3}
            &\# of Forks & E2, E4 \\
                & & The content is structured and independent. \\ 
                & & \url{https://help.github.com/en/articles/fork-a-repo} \\
                \cline{2-3}
            &\# of Open Issues & E2, E4 \\
                & & Structured and independent content. \\
                & & \url{https://help.github.com/en/articles/opening-an-issue-from-code} \\
            \hline
        \end{tabular}%
        }
        \label{tab:desc_of_feature}
    \end{center}
\end{table*}


Table \ref{tab:desc_of_feature} shows that channels with tacit forms of knowledge being externalized (i.e., externalized dimension of SECI).
Since the classification of tacit and explicit knowledge is not trivial, we applied the most distinguishable features taken from Table~\ref{tab:tacitandexplicitknowledge} (i.e. T2, T3, T4, E2, E3, and E4). 
In general we used the following rationale as guidance:
\begin{itemize}
	    \item \textit{T2 - Personal}: The knowledge possessed by any individual. Usually accumulated through observation or experiences. For the study, we characterize individual additions with no structure.
	    \item \textit{T3 - Context sensitive/specific:} The content is specific to its original context. It depends on particular time and space. Similar to T2, here the project customizes the channel specific to the project requirements or nature (i.e., library or framework, programming language)
	    \item \textit{T4 - Dynamically created:} The content is capable to change or customize. Since GitHub has templates, we regard these features are not in the templates.
	    \item \textit{E2 - Structured:} The information is organized in a predictable way and usually classified with metadata. For instance, a workflow tool usually has structure to is, when compared to a wiki.
	    \item \textit{E3 - Fixed content:} The content that is not, under normal circumstances, subject to change. This feature is more common with workflow and tools that serve as channels.
	    \item \textit{E4 - Context independent:} The content is unaffected by contextual relevance. For instance, the channel can serve different purpose for different projects.
	\end{itemize}

Interestingly, we find that the security audit is a mix of tacit and explicit forms.
Although the audit tends to personal, the contents that describe the strategy, policy and the process related to the management are fixed.
Thus, we conclude that the developers can provide the guidelines with regards to reducing the risk of misrepresentation of knowledge when developing software \cite{Lingzi:2015:7373782}.

\section{A Topological Analysis of Communication Channels Across GitHub Ecosystems}
\label{sec:TMS}
Taking the results from the preliminary study, we are now able to study the knowledge topology of these \cc{}.
This topology mapping analysis presents a visual representation of \cc{} within and across projects in the GitHub ecosystems.

\begin{figure}
    \centering
    \includegraphics[width=1\textwidth]{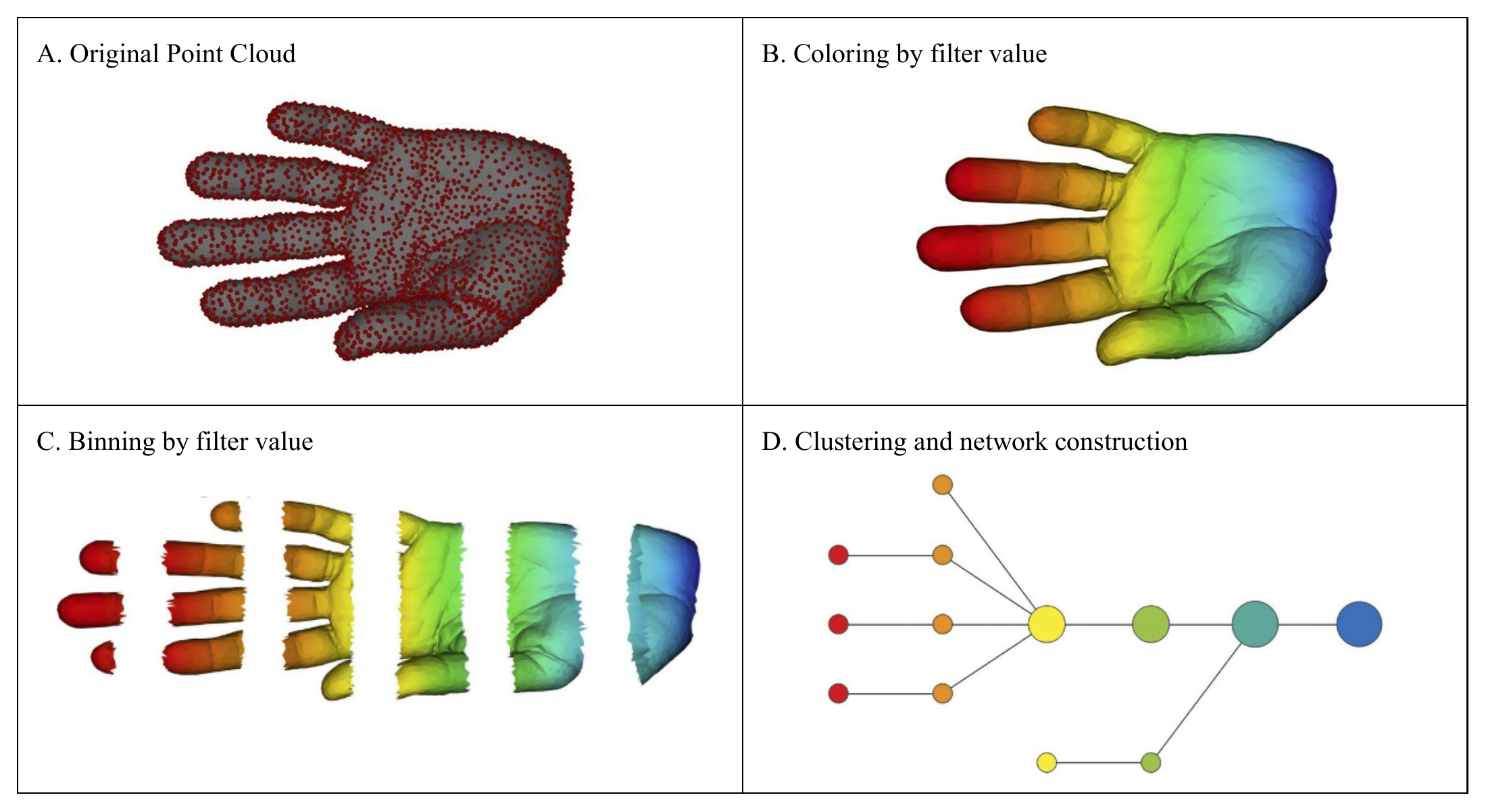}
    \caption{Taken from Lum et al.~\cite{Lum:2013:10.1038}, A) 3D object (hand) represented as a point cloud, B) A filter value is applied to the point cloud and the object is then colored by the values of the filter function, C) The dataset is binned by filter value, D) Each bin is clustered and a network is built.
    Within each cluster, groups of nodes determine the shape.}
    \label{fig:TDAhand}
\end{figure}

\subsection{Topological Data Analysis}
Due to the vast amount of data and the different communication channels, we apply the Topological Data Analysis (TDA) technique.
TDA is an approach to extract meaningful information from such data that is insensitive to the chosen metric, high-dimensional, noisy and incomplete without initiating a query or hypothesis~\cite{Lum:2013:10.1038}.
TDA has been employed in many research fields for data exploration and mapping.
Lum et al.~\cite{Lum:2013:10.1038} showed the significance of understanding the ``shape'' of data by implemented topology to analyze three different types of data: data of breast tumors to show gene expression, data of voting behavior from members of the United States House of Representatives and performance data of the NBA players.
In software engineering, TDA was also applied in a study of software testing by Costa et al.~\cite{PitaCosta:2017:TDA:3053600.3053604}.
Similar to Lertwittayatrai et al. \cite{Lertwittayatrai:2017:1710.00446}, a topology of the dataset is generated to provide a visual interpretation of multi-dimensions data analysis.

Figure~\ref{fig:TDAhand} provides an example of how a TDA is constructed.
TDA assumes a choice of a filter or its combination that can be viewed as a map to a space of metric to provides insights based on clustering the various subsets of the dataset related the choices of filter values. 
As shown, each node is represented as a set of data points, and the connection between nodes occurs if and only if their corresponding collections of data points have a point in common.
The topology is constructed by clusters of nodes (i.e., nodes connected together). 
Then within the cluster, we can find groups of nodes that form a shape of the dataset.
The density of the nodes and their shape gives an indication of the dominant of the features.
Tailored to our study, each point is a project that are clustered according to the different features extracted in the preliminary study. 
The use of color highlights the dominance of a feature, which indicates high occurrence of that channel.

For TDA, the clustering is performed using the t-Distributed Stochastic Neighbor Embedding (t-SNE)~\cite{Maaten2008}. 
In detail, the algorithm starts by calculating the probability of similarity of points in high-dimensional space, computing in proportion to their probability density under a Gaussian (normal distribution) algorithm.
Multi-dimensional data are then mapped by the t-SNE to a lower dimensional space and attempts to find patterns in the data by identifying observed clusters based on similarity of data points with multiple features.

\subsection{Motivation}
\label{sec:TMSmotivation}
Our motivation for the topological mapping study is to present a representation and overview of \cc{} that exists within large-scale ecosystems.
As such we formulated two research questions as follows:
\begin{itemize}
    \item \textit{\textbf{\RqTwo}}  \\
    In this research question, our motivation is to investigate how \cc{} evolve and change over time. 
    \\
    \item \textit{\textbf{\RqThree}}    \\
    For this, we take a closer look at the ecosystem. By studying the seven ecosystems, we are able to understand whether there are differences in knowledge.
\end{itemize}

\subsection{Approach}
Our approach to answer the two research questions is through the TDA mapping technique. 
The TDA mapper algorithm \cite{SPBG:SPBG07:091-100} uses combinatorial representations of geometric information about high-dimensional point cloud data, which is implemented with the Knotter tool \cite{SPBG:SPBG07:091-100}.
The tool provides a common framework which includes the notions of density clustering trees, disconnectivity graphs, and Reeb graphs, but which substantially generalizes all three. 
We use the t-Distributed Stochastic Neighbor Embedding (t-SNE) \cite{Maaten2008}, a technique for dimensionality reduction and clustering, and our defined features as the filters for the visualization construction.
For RQ$_1$, we analyze the map by identifying the most dense clusters (i.e., majority of projects) and then find the dominant features for those clusters.
For RQ$_2$, to find dominant features, we identify groups of nodes within the cluster and label.

\subsection{Data Collection}
As with the preliminary study, the same dataset from \texttt{libraries.io} was used in our experiments.
The results of the preliminary study in Table \ref{tab:desc_of_feature} were used as feature inputs in the topological mapper construction.
For RQ$_1$, we selected only projects between 2015 and 2017 because (i) they contained the youngest projects and (ii) all seven ecosystems had sufficient sample projects within this time period.
Part of the data preparation involved normalizing each of the 13 features into a value that ranges from 0 to 1. 
Based on the type column in Table \ref{tab:desc_of_feature}, we normalize the int, string and boolean values.
For integers, we calculate the ratio of the $X_{i,j,k}$ and $X_{i,j,max}$ which $X_{i,j,k}$ is the value of feature i in project k which is in platform j and $X_{i,j,max}$ is the maximum value of feature i in platform j. 
For boolean and string types, we represent 1 to them if the value is TRUE that indicates the channel exists.
On the other hand, represent as 0 if the project does not use that channel. 
The tool limited the maximum number of projects selected to 10,000, which resulted in selecting the top 10,000 most popular projects (based on the star count).

\subsection{Analysis}
Table \ref{tab:datasetevolutionRQ1} shows that on average we used up to 30,000 projects for each of the seven ecosystems from the libraries.io\footnote{dataset available at \url{https://libraries.io/}}.
Note that for $RQ_1$, we prepared an evolutionary set of topologies, dividing the dataset into three time periods (i.e., 2015, 2016 and 2017). 
Generation is approximated at up to 20--70 minutes for each of the 28 topologies.

\begin{table}[]
    \caption{Statistics of Generated Topologies including the Topology Build-time}
    \begin{center}
        \resizebox{\columnwidth}{!}{%
        \begin{tabular}{p{4.3cm}|r|r|r|r|r}
            \hline \hline
            \textbf{Library Ecosystem} & Pop. Size & RQ$_{1}$ \#proj. & RQ$_{1}$ \#proj. & RQ$_{1}$ \#proj.& RQ$_{2}$ \#proj.    \\
            & &  created 2015 & created 2016  & created 2017  &    \\
            \hline
            Go & 743,841 & 10,000 & 10,000 & 509 & 20,000  \\
            npm & 447,306 & 10,000 & 10,000 & 10,000 & 20,000  \\
            Packagist & 176,608 & 10,000 & 10,000 & 10,000 & 20,000    \\
            RubyGems & 93,377 & 10,000 & 10,000 & 8,611 & 20,000   \\
            PyPI & 69,895 & 10,000 & 10,000 & 10,000 & 20,000  \\
            Bower & 64,271 & 10,000 & 10,000 & 6,472 & 20,000  \\
            Maven & 62,654 & 10,000 & 7,526 & 428 & 20,000 \\
            \hline
            \textbf{build-time per} & & 35.03 &  29.60 & 20.22 & 70  \\
            \textbf{topology (mins.)} & & & & & \\
            \hline
            \hline
            \textbf{Totals} & 1,657,952 & 70,000 & 67,526 & 46,020 &  \\
            \hline
        \end{tabular}%
        }
    \label{tab:datasetevolutionRQ1}
    \end{center}
\end{table}

\underline{{Answering RQ$_{1}$}}:
To answer RQ$_{1}$, we split the projects to separate the older projects from the younger ones using the date that they were created (2015, 2016 or 2017).
Shown in Table \ref{tab:datasetevolutionRQ1}, we generate a topology that highlight the influencing features to explore differences between the older and younger projects. 
First, we identify main clusters of points. 
Then we compare these clusters over the three years.
Note that the color filter helps to identify dominant features.

\underline{{Answering RQ$_2$}}:
To answer RQ$_2$, we construct seven library specific topologies to find whether projects that are popular (i.e., has the most stars in that ecosystem) share similar channels across ecosystems.
First, we identify and analyze the dominant features of nodes.
Then, using the median score of stars per project within those nodes, we identify the group that contains more popular projects (i.e., labeled as Popular Group) when compared to the other groups (i.e., labeled as Non-Popular Group).

\begin{figure}[]
    \centering
    \begin{subfigure}{\textwidth}
        \center
        \includegraphics[width=0.95\textwidth]{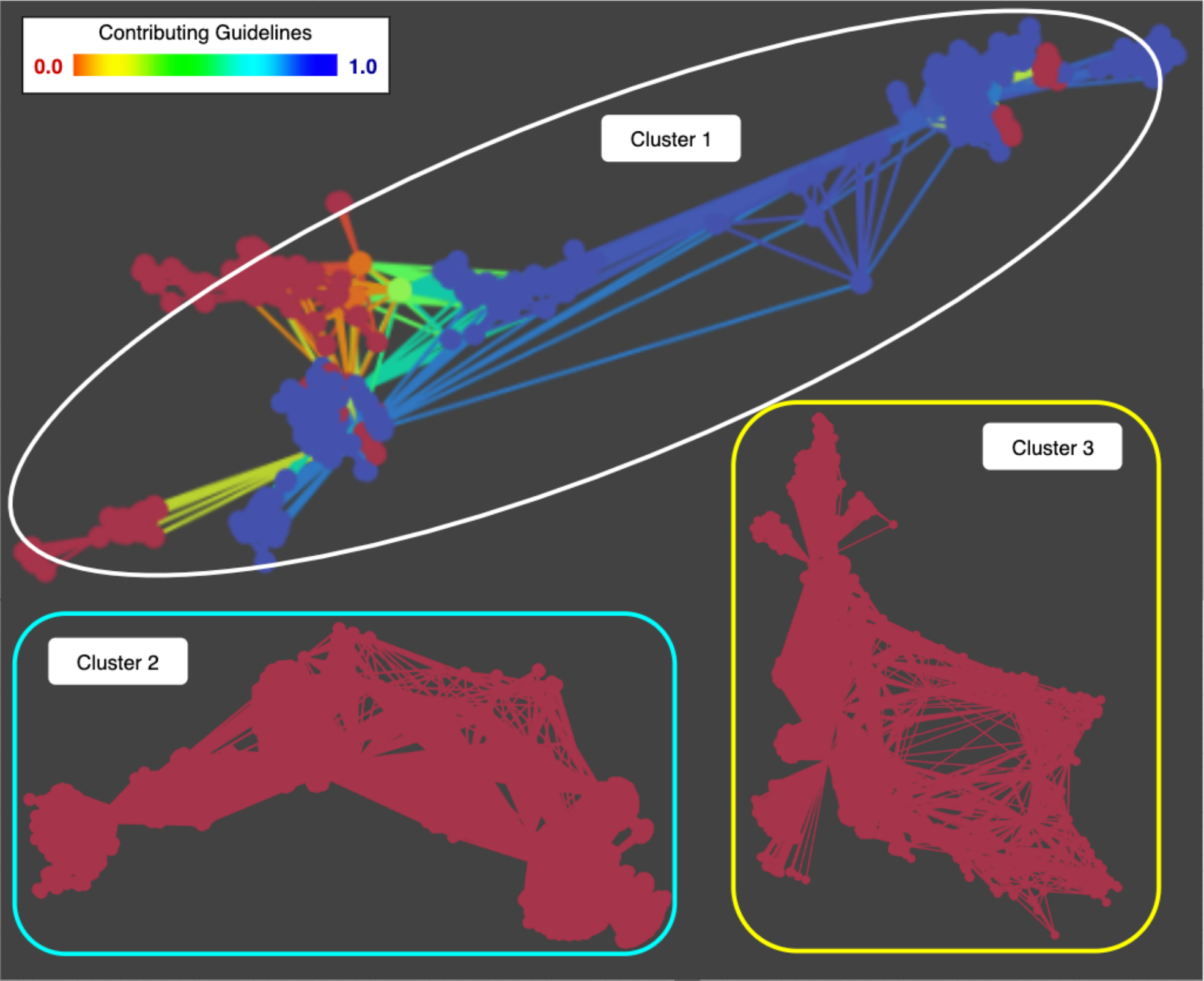}
        \caption{Projects created in 2015}
    \end{subfigure}
    \begin{subfigure}{0.465\textwidth}
        \center
        \includegraphics[keepaspectratio,scale=0.3,angle=0]{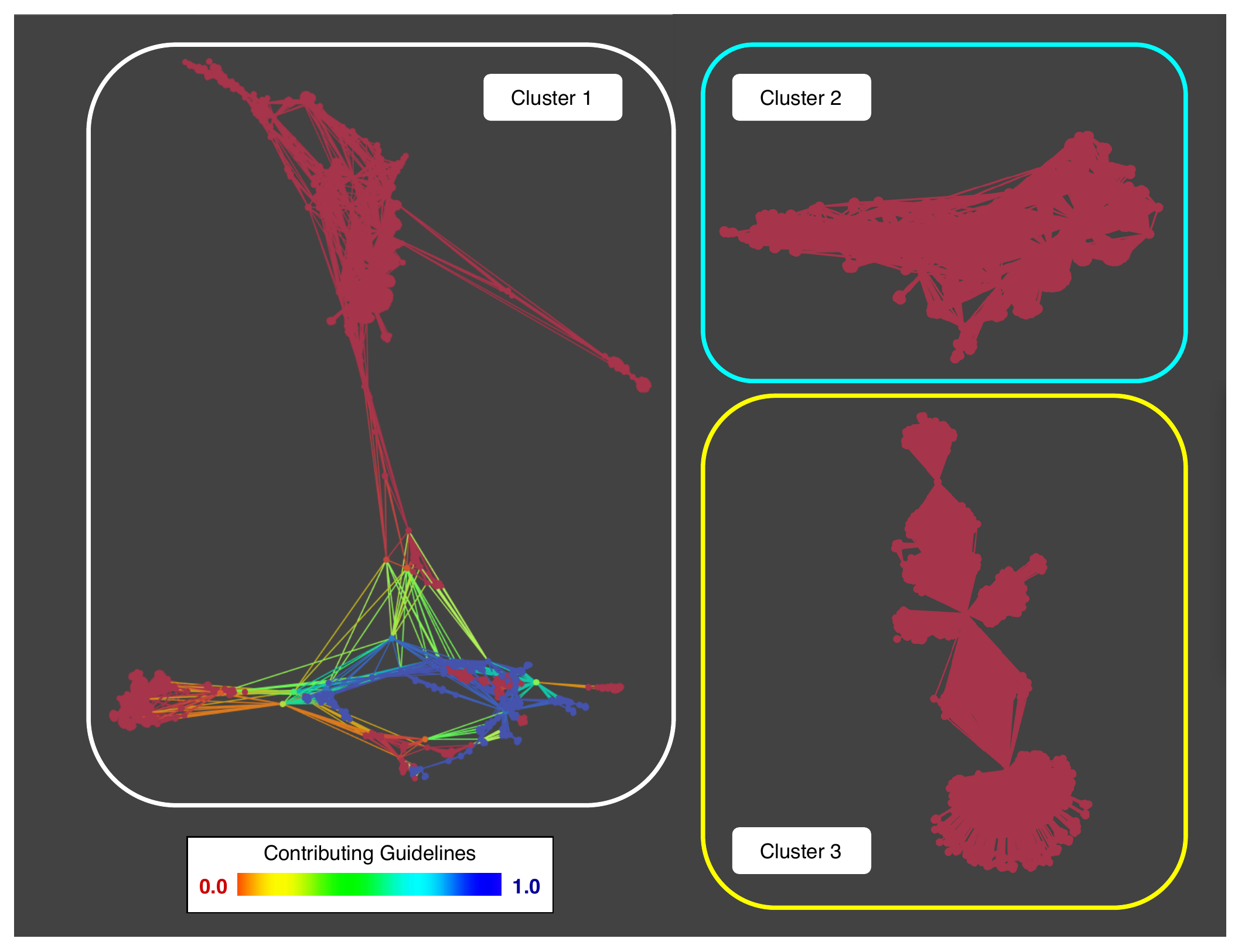}
        \caption{Projects created in 2016}
    \end{subfigure}
    \begin{subfigure}{0.465\textwidth}
        \center
        \includegraphics[keepaspectratio,scale=0.3,angle=0]{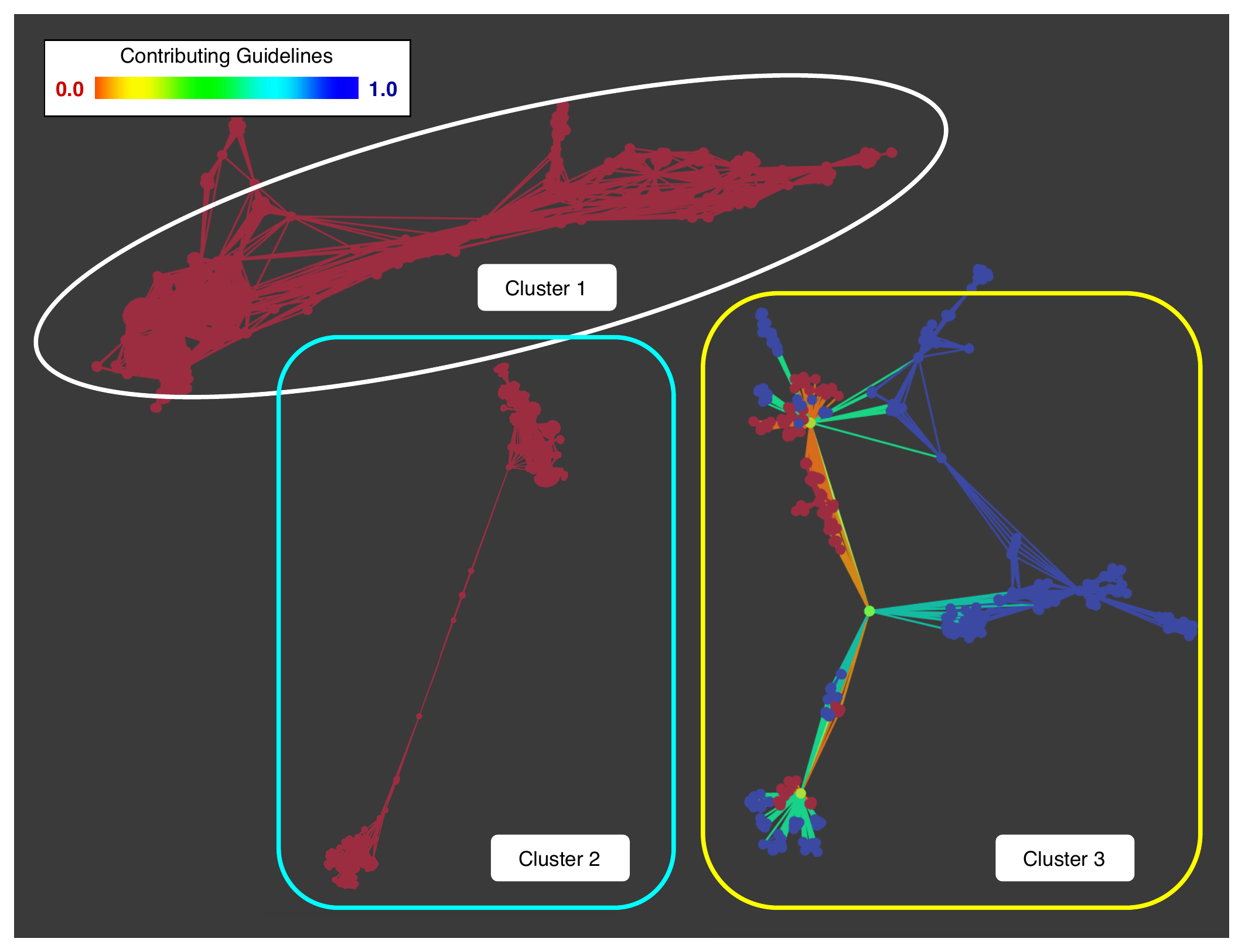}
        \caption{Projects created in 2017}
    \end{subfigure}
    \caption{Generated Topologies for projects created in (a) 2015, (b) 2016 and (c) 2017}
    \label{fig:RQ1}
\end{figure}

\subsection{Results}
\label{sec:results}
We now introduce our answer to the research questions and then describe the results.

\vspace{3mm}
\noindent \textbf{\RqTwo}
\begin{quote}
    \textit{`Younger projects adopt different channels compared to older projects'}
\end{quote}

We observed two main findings.
First, from Table \ref{tab:statisticresult}, the topology reveals that younger projects are adopting different \cc{} mechanisms when compared to the older projects.
To make the topology easier to read, we assigned the color (blue indicates existence while red indicates no existence) to the \texttt{Contributing} \texttt{Guidelines} feature.
Note that Cluster 1 always indicates the highest number of points (refer to Table~\ref{tab:statisticresult}).
Therefore, we can see that the blue nodes first are dominant in Cluster 1 in (i.e., Figure~\ref{fig:RQ1}(a)), but tend to become less dominant in 2016 and 2017 (i.e., Figure~\ref{fig:RQ1}(c)).

Second, as shown in the Table~\ref{tab:statisticresult}, we see that although some communication channels have changed over time (i.e. GitHub Pages, Security Audit, Changelog, Contributing Guidelines, and Fork), we also find that others (i.e. Wiki, Issue Tracker, and License) are still consistently used by most projects (Cluster 1).

\begin{table*}[]
    \centering
    \caption{Evolution of Externalization and Combination between 2015 and 2017}
    \label{tab:statisticresult}
    \resizebox{\textwidth}{!}{
    \begin{tabular}{lc|rr|aac|aaacc}
        \toprule
        \multirow{3}{*}{Period} & \multirow{3}{*}{Cluster} & \multirow{3}{*}{\#Nodes} & \multirow{3}{*}{\#Points} & \multicolumn{3}{c|}{Externalization} & \multicolumn{5}{c}{Combination} \\
        \cline{5-12}
        \noalign{\smallskip}
         & & & & \cellcolor{white} GitHub & \cellcolor{white} Security & \cellcolor{white} Wiki & \cellcolor{white} Changelog & \cellcolor{white} Contributing & \cellcolor{white} Fork & Issue & License  \\
         & & & & \cellcolor{white} Pages & \cellcolor{white} Audit & \cellcolor{white} & \cellcolor{white} & \cellcolor{white} Guidelines & \cellcolor{white} & \cellcolor{white} Tracker & \\
        \midrule
        \rowcolor{gray}
        2015 & 1 & 376 & 16,906 & \checkmark & - & \checkmark & \checkmark & \checkmark & \checkmark & \checkmark & \checkmark  \\
         & 2 & 4,943 & 15,638 & - & - & \checkmark & - & - & - & \checkmark & \checkmark  \\
         & 3 & 3,257 & 5,138 & \checkmark & - & \checkmark & - & - & - & \checkmark & \checkmark  \\
        \midrule
        \rowcolor{gray}
        2016 & 1 & 6,289 & 46,800 & \checkmark & \checkmark & \checkmark & \checkmark & \checkmark & \checkmark & \checkmark &\checkmark  \\
         & 2 & 1,377 & 7,650 & - & - & \checkmark & - & - & - & \checkmark & -  \\
         & 3 & 4,038 & 3,088 & \checkmark & - & \checkmark & - & - & - & \checkmark & \checkmark  \\
        \midrule
        \rowcolor{gray}
        2017 & 1 & 973 & 14,098 & - & - & \checkmark & - & - & - & \checkmark & \checkmark  \\
         & 2 & 1,595 & 8,794 & - & - & \checkmark & - & - & - & \checkmark & -  \\
         & 3 & 354 & 5,046 & \checkmark & - & \checkmark & \checkmark & \checkmark & \checkmark & \checkmark & \checkmark  \\
        \bottomrule
    \end{tabular}
    }
\end{table*}

\vspace{3mm}
\noindent \textbf{\RqThree}

\begin{quote}
    \textit{`Library ecosystems employ channels that capture new knowledge (i.e.,  externalization). Channels updating existing knowledge (i.e., Combination knowledge) varies from one ecosystem to another'}
\end{quote}

\begin{figure}[]
    \centering
    \begin{subfigure}{\textwidth}
        \center
        \includegraphics[width=0.95\textwidth]{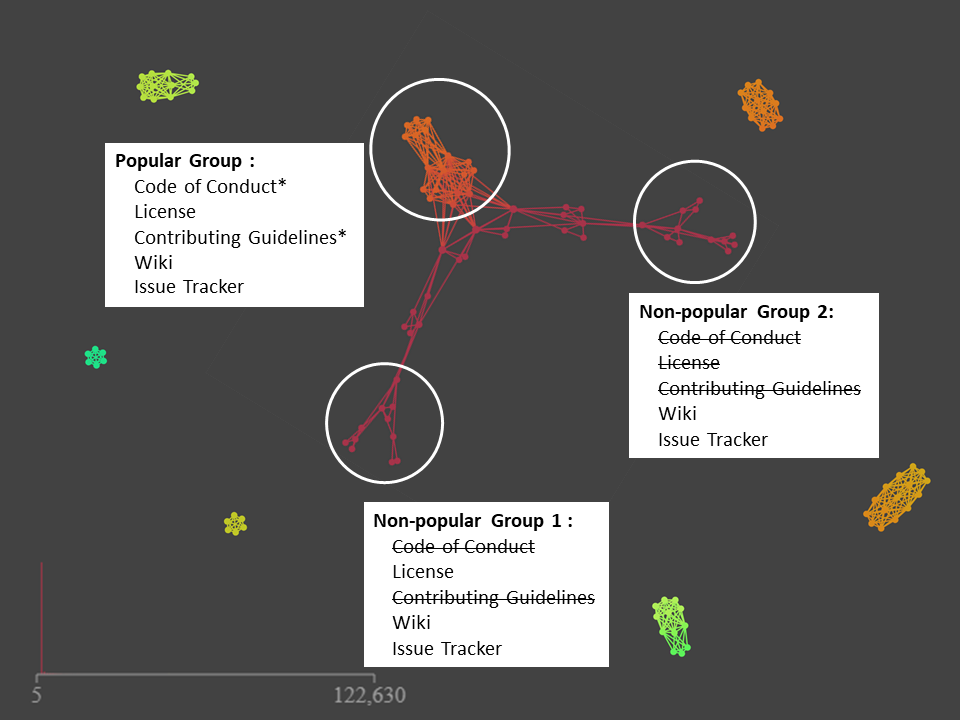}
    	\caption{Topology for Bower libraries}
    \end{subfigure}
    \begin{subfigure}{0.465\textwidth}
        \center
        \includegraphics[keepaspectratio,scale=0.25,angle=0]{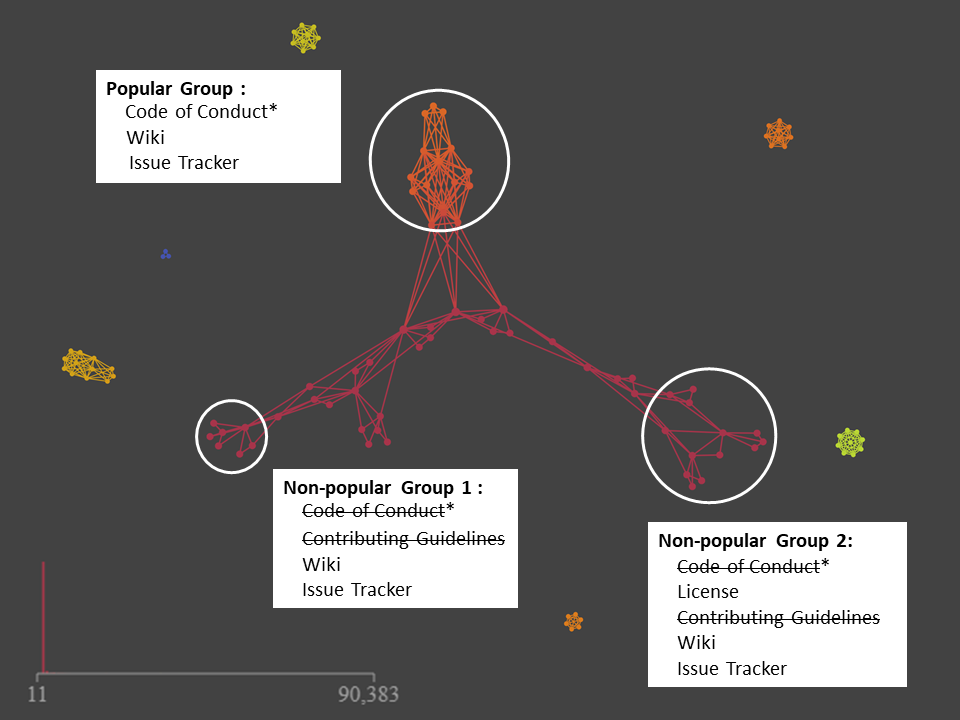}
    	\caption{Topology for PyPI libraries}
    \end{subfigure}
    \begin{subfigure}{0.465\textwidth}
        \center
        \includegraphics[keepaspectratio,scale=0.25,angle=0]{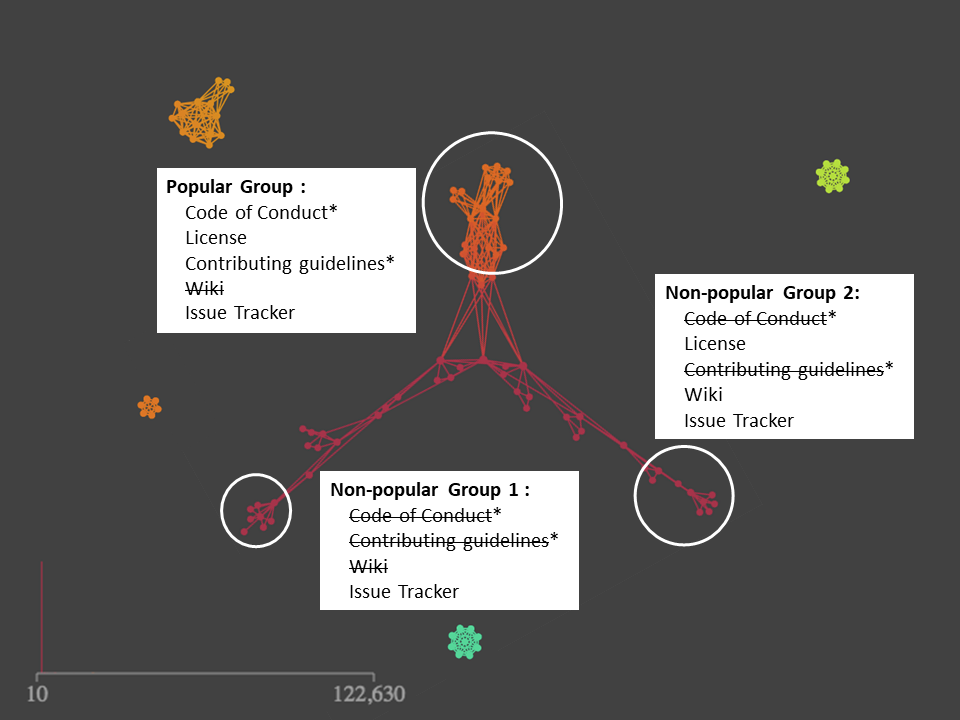}
    	\caption{Topology for RubyGems libraries}
    \end{subfigure}
    \caption{Topology for three of the seven ecosystems (a) Bower, (b) PyPI, and (c) RubyGems}
    \label{fig:RQ2}
\end{figure}

In terms of the topology shape, Figure~\ref{fig:RQ2} depicts ecosystems (i.e. Bower, PyPI and RubyGems) having triangular shape topology with three main group points.
This is consistent for the rest of the studied ecosystems.
Under further investigation, we see that one of the group represents the popular projects (i.e., popular), while the other two group points were the non-popular data points (i.e., non-popular 1 and 2).

Table \ref{tab:featuresacrosslibraryecosystems} shows two results.
For both popular and non-popular projects, we find that the issue tracker has been a consistent communication channel for applying explicit knowledge (i.e., combination).
Combining with the results of $RQ_1$, one explanation could be that these are older projects.
Second, each ecosystem depicts a different set of explicit dominant features. 
For example, the Bower ecosystem includes combination knowledge transfer forms (i.e., Code of Conduct, License, Contributing Guidelines, Wiki, Issue Tracker), while PyPI projects are less likely to include a license or contributing guidelines. 
One reason could be that the license information is embedded in other locations, such as a webpage.
For example, the python library scikit-learn has its license information on the python ecosystem website\footnote{license at \url{https://pypi.org/project/scikit-learn/}}.

\newcolumntype{g}{>{\columncolor{gray}}c}
\begin{table}[t]
    \caption{Dominant Extracted Features Topologies across the Ecosystems}
    \label{tab:featuresacrosslibraryecosystems}
    \centering
    \resizebox{\columnwidth}{!}{%
    \begin{tabular}{l|l|l|g|g|c|c|g|c|c}
        \hline
        \textbf{Topology Cluster} & \textbf{Features} & \textbf{Dimensions} & \textbf{Bower} & \textbf{PyPI} & \textbf{Go} & \textbf{npm} & \textbf{RubyGems} & \textbf{Packagist} & \textbf{Maven} \\
        \hline
        Popular & Code of Conduct & Combination & \checkmark & \checkmark & - & - & \checkmark & - & -   \\
        & Contributing Guidelines & Combination & \checkmark & - & \checkmark & \checkmark & \checkmark & \checkmark & -    \\
        \rowcolor{gray}
        \cellcolor{white} & Issue Tracker & Combination & \checkmark & \checkmark & \checkmark & \checkmark & \checkmark & \checkmark & \checkmark \\
        & License & Combination & \checkmark & - & \checkmark & \checkmark & \checkmark & \checkmark & \checkmark  \\
        & Wiki & Externalization & \checkmark & \checkmark & - & - & - & - & \checkmark  \\
        \hline
        Non-popular & Code of Conduct & Combination & - & - & \checkmark\checkmark & - & - & - & -  \\
        & Contributing Guidelines & Combination & - & - & - & - & - & - & - \\
        \rowcolor{gray}
        \cellcolor{white} & Issue Tracker & Combination & \checkmark\checkmark & \checkmark\checkmark & \checkmark & \checkmark\checkmark & \checkmark\checkmark & \checkmark\checkmark & \checkmark \\
        & License & Combination & \checkmark & \checkmark & \checkmark & \checkmark & - & \checkmark & \checkmark  \\
        & Wiki & Externalization & \checkmark\checkmark & \checkmark\checkmark & \checkmark & - & \checkmark & \checkmark\checkmark & -   \\
        \hline
    \end{tabular}%
    }
\end{table}

Our study results also confirm that issue tracker as an important communication channel is common for both popular and non popular projects, as described in Figure~\ref{fig:RQ2} and Table~\ref{tab:featuresacrosslibraryecosystems}.
Issue trackers serve not only as part of the workflow and process (code maintenance and evolution) for software development, but also plays a significant role of communication in a software development process that store a large amount of data, such as discussion during triage meetings, reproduction step clarifications between the person who created an issue and its owner, etc~\citep{Bertram:2010:CCB:1718918.1718972}.
Other work such as Dings{\o}yr and R{\o}yrvik~\cite{Dingsoyr:2003:ESI:776816.776827} confirms that issue tracker builds up a substantial amount of information concerning the issue reports from customers, partially complete feature ideas, and the communication surrounding the software development. 
This large amount of information is often beneficial for both the organization and the software project team on a number of different levels.

\section{Topology Evaluation}
\label{sec:evaluation}

In comparison with other analysis methods, TDA has been proven to deal with both small and large scale patterns that often other techniques fail to detect.
Other more traditional method of analysis is the principal component analysis (PCA), multidimensional scaling (MDS), and cluster analysis.
Unlike traditional statistical methods, TDA does not provide any statistical test that is performed to support the observation. 
To evaluate and validate our use of TDA, we compared our method to the PCA method.
The main different with PCA is that it simplifies the complexity in high-dimensional data by transforming the data into fewer dimensions (i.e., usually into a x and y axis, depicted by a scatterplot), which act as summaries of features. 

\begin{figure}[]
    \centerline{\includegraphics[width=0.7\textwidth]{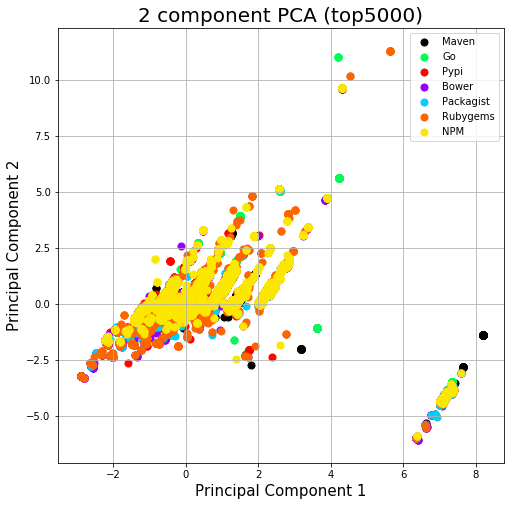}}
    \caption{A replication of RQ$_{1}$ using PCA. PCA does show location of the ecosystems of different platforms, but since the features are combined, we cannot identify the dominant channels.}
    \label{fig:TDA}
\end{figure}

In our approach, we apply the PCA method using the \texttt{sklearn.decomposition} python library\footnote{documentation at \url{https://scikit-learn.org/stable/modules/generated/sklearn.decomposition.PCA.html}} to visually examine the results.
For the evaluation, we will regenerate the results for RQ$_2$ and determine if we can visually identify dominant features within each ecosystem.

Figure~\ref{fig:TDA} shows the results of evaluating the TDA technique against the statistical Principle Component Analysis (i.e., PCA) method.
The PCA method shows the location of the ecosystems of different platforms and is able to summarize the features into two principle components.
However, the analysis is unable to show details of each feature, outlining (i) which features are dominant and (ii) how the features are different to each other.

\section{Implications}
\label{sec:implications}
Based on our results, we discuss three implications of the results in relation to the nature of communication channels for both researchers and practitioners.

\begin{enumerate}
    \item \textit{Contemporary GitHub Projects will continue to adopt multiple Communication Channels}. 
    Results indicate that GitHub projects are adopting 13 communication channels. 
    Thus the topological mapping is able to cluster together projects with similar channels. As shown in the topological evaluation, other techniques is able to map these relationships.
    The implication for researchers and practitioners is that knowledge is not stored in one channel, thus multiple channels must be considered to fully understand the knowledge shared in software projects.
    
	\item \textit{Communication Channels will change and evolve over time.}
	Results indicate that communications are constantly changing.
	For instance in RQ$_{1}$, channels like contributing guidelines have changed over time compared with the consistent ones like issue tracker.
    Furthermore, in RQ$_{2}$, we find that there are differences between popular and non-popular projects in different ecosystems.
    Results indicate that some channels are ecosystem specific.
    For example, contributing guidelines are commonly used by seven targeted ecosystems, except PyPI and Maven. While PyPI and Maven use wiki pages alongside Bower where this channel is not prevalence for the other ecosystems. This means the higher starred projects tend to move from externalization to combination.
    The implication for researchers and practitioners is that understanding the popular channels will help understand where knowledge is shared. For instance, we envision researchers should keep up with the newer channels to understand knowledge sharing within younger projects.

    \item \textit{Knowledge in Communication Channels is both external and combination}. Communication channels are used to capture new knowledge (i.e., externalization knowledge). 
    For example, the externalization of the Wiki is very popular.
    As shown in the preliminary study, the Wiki has some tacit features of being personalized to match the individual project.
    \\
    In contrast, updating existing knowledge in communication channels is common. 
    As mentioned by GitHub, contributing guidelines \textit{help them (developers) verify that they're submitting well-formed pull requests and opening useful issues}\footnote{\url{https://help.github.com/en/articles/setting-guidelines-for-repository-contributors}}. 
    GitHub projects are also encouraged to use the platform workflow, with the Issue Tracker becoming a popular tool and communication channel for developers.
    The final example is the License channel. 
    Putting a license has become increasingly important, especially for projects intended for library reuse. 
    This practice may also be community driven. 
    For example, according to Lertwittayatrai et al.~\cite{Lertwittayatrai:2017:1710.00446}, npm projects tend to use the MIT license in their projects.
    The implication for researchers and practitioners is that understanding where new knowledge is shared. This information could be very useful, for instance, especially for newbies to a project.
\end{enumerate}

\section{Threats to Validity}
\label{sec:threats_to_validity}

We discuss three key threats to the validity of the study.
The first relates to the categorization of knowledge. 
Nonaka and Takeuchi's categorization has been contested in CSCW~\cite{Schmidt:2012:TTK:2205458.2205469}, especially in terms of the tacit knowledge. 
By adopting the SECI model, we identify channels that have a possibility to capture tacit knowledge.
Furthermore, we focus on channels and how they are  important for project attractiveness and sustainability.

The second threat is related to the experiment setup and methodology.
In this work, we extract common collaborative channels as shown in prior works \cite{Borges:7816479,Anvik:2005:COB:1117696.1117704,Hauff:2015:MGD:2820518.2820563,Vendome:2015:7181450,Burrow:2004:NAW:1012807.1012831}, providing confidence in our channel selection.
To reduce feature bias, we used a normalized score in formulating features for the TDA. 
One key threat is the quality of the channel. For example, the existence of readme files in a project doesn't mean it is used or contains valuable information.
This is outside the current scope of work, however, future investigations will focus on quality of these channels and how much knowledge they contain. 
Finally, we use the star count rating in our project selection, yet this metric has been related to skewedness and not being normalized.
Since we assume that social coding is related to the social sharing nature of communication channels, we believe our use of star count is a useful proxy of projects that are more likely to actively use communication channels.

The third threats to validity are the accuracy and the limitation of the tools, especially whether results will change according different sample sizes.
As such, we use the largest sample of 10,000 points to ensure confidence in our result. As shown by Lertwittayatrai et al.~\cite{Lertwittayatrai:2017:1710.00446}, the result tends to stabilize as more points are added.

\section{Related Work}
\label{sec:developer_knowledge_sharing}
In this section, we present related works that complement this study organized into these (i) Communication Channels, (ii) Sharing Architectural Knowledge and (iii) the use of Topological Data Analysis.

\subsection{Communication Channels}

Several studies in other fields analyzed channels as the exchange of information.
In organization management, communication methods, whether  verbal or nonverbal messages to produce meanings in heterogeneous contexts, cultures and media~\cite{keyton-032516-113341}.
Channels are practical in a complex network of relationships where messages are created, delivered and received by individuals, as well as other communication practices that allow larger democracy \cite{keyton-032516-113341}. 
A study by Wang et al.~\cite{WANG2019135} investigated the usability, purposes and challenges of \cc{} in industry during safety analysis.
Related to software development, communication between developers is possible to augment through collaborative programming~\cite{Williams:2002:PPI:548833}, or direct communication between team members~\cite{Schwaber:2001:ASD:559553}.
Therefore, the \cc{} design or the necessary of social skills in organization management receive more attention from researchers.
Lindsj{\o}rn et al.~\cite{LINDSJORN2016274} analyzed communication technique to measure the teamwork quality in influencing the performance of software teams and the successful of their team members.
The finding indicates that the quality of teamwork in agile teams does not tend to be higher than traditional teams in other similar survey.
Team performance is the only effect of teamwork quality which is greater for agile teams than traditional teams.
Our study complements these studies, showing how communication channels are indicators of knowledge in a software organization.

In the field of Software Engineering, research into channels is based on social practices.
Social practice characterizes the existence of activities which are related to each other \cite{DITTRICH2016220}. 
These collaborative works are conducted through (i) distributed teleo-affective structures for software design and development, (ii) shared common or specific knowledge of the software development requirements, and (iii) clear procedures and regulations governing people to accomplish specific activities\footnote{\label{socialpracticedef}\url{https://en.wikipedia.org/wiki/Practice_theory}}.
Social practices are not confined to only industry-related practices, but more broadly, they can be implemented in open source software projects. 
In addition to requiring a shared understanding of the requirements to become a project member, a development of open source products performed by complying with common rules as well, and by using a shared teleo-affective.
Therefore, the activities of each individual can be connected from the initial of the development to the end of the project.
The example of the requirements in the open source projects is also described by Scacchi~\cite{Scacchi:999088}. 
This study analyzes channels from a knowledge perspective instead of social collaborations.
Our topology confirms that the collaborative and participatory nature of software development continues to evolve, shape, and be shaped by communication channels that are used by development-related communities of practice~\cite{lanubile2013social}.
A study undertaken by Treude and Storey~\cite{Treude:2011:ECS:2025113.2025129} shows that different media artifacts and channels used for knowledge sharing have different implications for software development.
We believe that the methods such as ecosystem topology can provide us a more empirical means to assess inconspicuous patterns within an ecosystem.
For example, the topology can reveal the type of channels that were used by most projects.

There is related work that specifically studied GitHub projects, especially library ecosystems and their social collaborations.
The social features used in a social coding platform, such as GitHub, has attracted many researchers to analyze. 
The collaborative features used in their studies, including open bug repositories \cite{Anvik:2005:COB:1117696.1117704}, project fork \cite{Borges:7816479,Hauff:2015:MGD:2820518.2820563}, the usage of a software license \cite{Vendome:2015:7181450,Wu:2017}, and the use of wiki \cite{Burrow:2004:NAW:1012807.1012831}. 
Open bug repositories, such as the Bugzilla\footnote{\label{bugzillaweb}\url{https://www.bugzilla.org/} (July 2018)}, are mostly managed by open source projects to allow users to be more contributing. 
Anvik et al.~\cite{Anvik:2005:COB:1117696.1117704} stated that even though these repositories are often used as a reference by open source developers, however the data availability on how they interact with the issues tracking systems is limited. 
A work carried out by Borges et al.~\cite{Borges:7816479} studied that the popularity of a project on GitHub relies on some factors such as the language that developers used to program and the domain of the application. 
These main elements were presumed to impact on the number of stars of a project. 
A prior study also analyzed the evolution of software licenses empirically \cite{Vendome:2015:7181450}. 
To complement prior work, this work looks at all channels to provide a holistic viewpoint of all the different channels.

\subsection{Sharing Architectural Knowledge}
The impact of communication channels in Sharing Architectural Knowledge has been highlighted in several studies outside of Software Engineering.
For instance, Borrego et al.~\cite{Borrego:7577427} conducted an empirical study to investigate agile methodologies articulation in unstructured and textual electronic media (such as emails, forums, chats etc.) in global software development.
The findings show the involvement of aspects in architectural knowledge in the unstructured and textual electronic media in the teams.
Architectural knowledge in the unstructured media is also perceived as important, regardless the interaction frequency.

In a software engineering context, other work studied how knowledge in communication channels impact project and their ecosystem success.
Failing FLOSS projects provide insights into some of the outside forces that detract developers from making contributions.
A study by Coelho et al.~\cite{Coelho:FSE2017} found the following reasons for failing projects: usurped by competitor, obsolete project, lack of time and interest,  outdated technologies, low maintainability, conflicts among developers, legal problems, and acquisition.
To mitigate these reasons, projects need to attract as well as retain its existing base of contributors.
In fact, Hata et al.~\cite{HataChase15} suggests that improving the code writing mechanisms (i.e., wikis, official webpage, contributing and coding guidelines and using multi-language formats) leads to more sustainable projects.
A study by Storey et al.~\cite{Storey17} showed that ecosystems of FLOSS projects are shaped through social and communication channels (sometimes referred to as social coding).
Recently, Aniche et al.~\cite{Aniche:2018} confirmed that news channels also play an important role in shaping and sharing knowledge among developers.
Hence, owners of projects could boost their social presence through participation on recent topics from news aggregators such as \texttt{reddit}\footnote{\url{https://www.reddit.com}}, \texttt{Hacker News}\footnote{\url{https://news.ycombinator.com}} and \texttt{slashdot}\footnote{\url{https://slashdot.org}}.
In addition, a study conducted by Tamburri et al.~\cite{Tamburri2018} described that the characteristics of ecosystem measurement can also be utilized to explain the structure of open-source ecosystem pattern.
Our results complement these work and have the similar goal of understanding how projects can attract developer contributions.

\subsection{Topological Data Analysis (TDA)}
The TDA technique has been applied in different research fields outside of software engineering.
For instance, a study by Lum et al.~\cite{Lum:2013:10.1038} used TDA to investigate three different cases, namely, patient identification in breast cancer, implicit networks of the US House of Representatives, and NBA team stratification.
The study shows that TDA can handle various types and high-dimensional datasets using three real world examples. 
From the analysis, the TDA shows the shapes of the breast cancer gene expression networks that allow to identify subtle but potentially biologically relevant subgroups, the shapes of the networks formed across the years about the voting patterns of the members of The US House of Representatives, and the playing styles of the NBA players.

In the software engineering context, the TDA topology has also been applied in such studies.
Lertwittayatrai et al.~\cite{Lertwittayatrai:2017:1710.00446} use topological methods to visualize the high-dimensional datasets from a software ecosystem.
In the study, the TDA allows the analysis of relationships between six related dataset features of a package, that is, author, author domain, license, tagged keywords, version released, and number of dependencies.
In our work, we combine all communication channels to understand at a higher level how projects in the ecosystem use communication channels to capture and share knowledge.

\section{Conclusion}
\label{sec:conclusion}

To understand what knowledge sharing occurs in communication channels, we conducted an analysis of channels in 70 thousand GitHub projects. 
First we conducted a preliminary study to identify and map what knowledge exists and is transferred through the 14 channels. 
We then used the topological mapper to provide a high-dimensional visual shape of the communications over time and for different library ecosystems.
Our work shows that GitHub projects tend to adopt multiple channels.
Furthermore, these channels changing over time and can be classified as either capturing new knowledge or updating the existing knowledge.

Based on this work, which established the role of multiple communication channels with knowledge sharing, there are many open avenues for future work: understanding the role and the different combination usage of channels, further studies into cross-channel knowledge, and tool support for channel recommendations, to name a few.

\section*{Acknowledgement}
This work has been supported by JSPS KAKENHI Grant Number 16H05857, 17H00731, and 18H04094.

\section*{References}


\end{document}